\documentclass[onecolumn,sort&compress,numbers]{els-mrw} 

\usepackage{amsmath,amssymb,amsfonts,amsthm,makeidx,graphicx}
\usepackage{txfonts}
\usepackage{helvet}

\begin{document}


\chapter{An introduction to effective potential methods in field theory
}\label{chap1}

\author[1]{Isabella Masina}%
\author[2]{Mariano Quiros}%


\address[1]{\orgname{University of Ferrara and INFN}, \orgdiv{Dept. of Physics and Earth Science}, \orgaddress{Via Saragat 1, 44122 Ferrara, Italy}}
\address[2]{\orgname{Institut de Fisica d'Altes Energies (IFAE)}, \orgdiv{and The Barcelona Institute of Science and Technology (BIST)}, \orgaddress{Campus UAB, 08193 Bellaterra, Barcelona Spain}}

\articletag{Chapter Article tagline: update of previous edition, reprint.}

\maketitle

\begin{abstract}[Abstract]
	In this chapter we give a pedagogical introduction to effective potential methods in field theories. We first review the general functional methods leading to the concept of effective action and effective potential. Focusing on the effective potential we review the methods to compute radiative contributions, starting from the original 1973 seminal paper by Sidney Coleman and Erick Weinberg on one-loop corrections, along with its improvement by the renormalization group equations, as well as some basic results from thermal field theory, including the one-loop thermal corrections to the effective potential. We concentrate on some physical applications of the effective potential: \textit{i)} the possibility of spontaneous symmetry breaking by radiative corrections, \textit{ii)} the metastability of the electroweak vacuum in the Standard Model, and \textit{iii)} the relevance of quantum effects in the dynamics of Higgs inflation. We then apply the previous results to cosmological phase transitions in the early universe and the possibility of generating a stochastic background of gravitational waves as well as primordial black holes.    
\end{abstract}

\begin{keywords}
 	effective potential\sep thermal corrections \sep electroweak metastability\sep phase transitions\sep Higgs inflation 
	
\end{keywords}



\section*{Objectives}
\begin{itemize}
	\item In Sec.~\ref{sec:effective} we provide basic results on the effective potential at the loop level, in theories at zero and finite temperature. 
	\item In Secs.~\ref{sec:radiative}-\ref{sec:quantum},  we study some relevant applications at zero temperature of the previous results: in particular the possibility of radiative breaking and the relevance of radiative corrections for establishing the metastability of the electroweak minimum, at very high scales, and for inflationary models, in particular for Higgs inflation.
	\item In Sec.~\ref{sec:cosmological} we study the application of previous results on thermal field theory to phase transitions, and the generation of a stochastic gravitational waves background and production of primordial black holes in first order phase transitions.
\end{itemize}


\section{Introduction}\label{intro}

The effective potential in quantum field theory (QFT) is a useful tool to compute the vacuum (or ground) state (i.e.~the state of minimum energy) of a given theory in particle physics. It was originally introduced by Euler, Heisenberg and Schwinger~\cite{Heisenberg:1936nmg,Schwinger:1951nm}, and applied to studies of spontaneous symmetry breaking by Goldstone, Salam, Weinberg and Jona-Lasinio~\cite{Goldstone:1962es,Jona-Lasinio:1964zvf}. The effective potential, at the lowest order, is simply given by the tree-level potential which is in general a function of the scalar field whose background value is at the origin of the symmetry breaking. Its most celebrated example is the potential of the Standard Model (SM) for which the scalar field is the Higgs boson, discovered at the Large Hadron Collider (LHC) at CERN in 2012 by the ATLAS and CMS experiments~\cite{ATLAS:2012yve,CMS:2012qbp}. In theories beyond the Standard Model the scalar fields, upon which the effective potential depends on, can have more complicated structures: e.g.~two scalar fields, in the case of electroweak theories with extended Higgs sector (as the two Higgs doublet model, or the minimal supersymmetric Standard Model, with two Higgs doublets), or even multi-scalar fields as the case of grand unified theories spontaneously broken to the Standard Model by scalars in the adjoint representation of the gauge group, as e.g.~SU(5) or SO(10).

Of course the tree-level potential is just the classical approximation which has to be completed, in precision calculation, by loop-effects.  Explicit calculations of the effective potential including radiative corrections were initially performed at the one-loop level by Sidney Coleman and Erick Weinberg~\cite{Coleman:1973jx}, and at higher-loop by Jackiw~\cite{Jackiw:1974cv}, Iliopoulos, Itzykson and Martin~\cite{Iliopoulos:1974ur}, and more recently by many other authors, see e.g.~two~\cite{Martin:2001vx}, three~\cite{Martin:2015eia} and four-loop~\cite{Martin:2013gka} results. In this chapter we will provide a pedagogical introduction to the effective potential methods used in the literature and some of their most interesting physical applications.

\section{The effective potential}
\label{sec:effective}

In this section we will provide the formalism for computing the effective-potential at different orders in perturbation theory, first at zero and then at finite temperature. Particularly interesting will be the improvement of the effective potential by the renormalization group equations (RGE), in particular using dimensional regularization and the so-called $\overline{\rm MS}$ renormalization scheme, which will introduce a non-physical scale $\mu$ in the theory. More details can be found in the lecture notes~\cite{Quiros:1994dr,Quiros:1999jp,Quiros:2007zz} by one of the authors.

\subsection{Effective potential at zero temperature}

We will consider the theory described by a scalar field $\phi$ with a Lagrangian density $\mathcal L[\phi(x)]$ and an action defined by the functional $S[\phi]=\int d^4x \mathcal L[\phi(x)]$. For an external current $j$ coupled to the field $\phi$ as $\phi\cdot  j\equiv \int d^4x \phi(x)j(x)$, different functionals are then defined: the generating functional (vacuum-to-vacuum amplitude) $Z[j]$, the connected generating functional $W[j]$, and the effective action $\Gamma[\bar\phi]$, as the Legendre transform of $W[j]$:
\begin{equation}
Z[j]\equiv\int d\phi e^{i(S[\phi]+j\phi)}\equiv e^{iW[j]},\quad \Gamma[\bar\phi]=W[j]-\int d^4x\, \bar\phi(x)j(x),\quad \textrm{where}\quad \bar\phi(x)\equiv\delta W[j]/\delta j(x) \,\,.
\end{equation}

The functionals $Z[j]$ and $W[j]$ can be expanded in a power series of the current $j$ as
\begin{equation}
Z[j]=\sum_{n=0}^\infty\frac{i^n}{n!}\int d^4x_1\dots d^4x_n j(x_1)\dots j(x_n)G^{(n)}(x_1,\dots x_n),\quad i\,W[j]=\sum_{n=0}^\infty\frac{i^n}{n!}\int d^4x_1\dots d^4x_n j(x_1)\dots j(x_n)G^{(n)}_c(x_1,\dots x_n)\,\,,
\end{equation}
the coefficients being the Green $G^{(n)}$ and the connected $G^{(n)}_c$ Green functions, respectively. Similarly the effective action can be expanded in powers of $\bar\phi(x)$ and its Fourier transform $\tilde\phi(p)=\int d^4x \exp(-i px)\bar\phi(x)$
\begin{equation}
\Gamma[\bar\phi]=\sum_{n=0}^\infty\frac{1}{n!}\int d^4x_1\dots d^4x_n \bar\phi(x_1)\dots \bar\phi(x_n)\Gamma^{(n)}(x_1,\dots x_n)=\sum_{n=0}^\infty \frac{1}{n!}\int\prod_{i=1}^n\left[\frac{d^4p_i}{(2\pi)^4} \tilde\phi(-p_i)\right]
(2\pi)^4\delta^{(4)}(p_1+\dots +p_n)\tilde\Gamma^{(n)}(p_1,\dots,p_n)\,\,,
\end{equation}
the coefficients $\Gamma^{(n)}$ and $\tilde\Gamma^{(n)}$ being the one-particle irreducible (1PI) Green functions in configuration and momentum space, respectively. 

In a translationally invariant theory where the $\bar\phi$ configuration is constant, i.e.~$\bar\phi(x)=\phi_c$ and consequently $\tilde\phi(p)=(2\pi)^4 \phi_c \delta^{(4)}(p)$, the effective action and the effective potential $V_{\rm eff}(\phi_c)$ can be written as
\begin{equation}
\Gamma(\phi_c)=\sum_{n=0}^\infty \frac{1}{n!}\phi_c^n (2\pi)^4 \delta^{(4)}(0)\tilde\Gamma^{(n)}(0)\equiv -\int d^4x V_{\rm eff}(\phi_c)\quad \Longrightarrow\quad V_{\rm eff}(\phi_c)=-\sum_{n=0}^\infty\frac{1}{n!}\phi_c^n \tilde\Gamma^{(n)}(0)\,.
\label{eq:effpot}
\end{equation}

The last expression in (\ref{eq:effpot}) is used for explicit calculations of the effective potential at any loop $n$ in perturbation theory such that the effective potential can be expanded as
\begin{equation}
V_{\rm eff}(\phi_c)=\sum_{n=0}^\infty V_n(\phi_c)=V_0(\phi_c)+V_1(\phi_c)+\dots \, \, ,
\end{equation}
where $V_0$ is the tree level potential which, typically, we can write as a degree four polynomial (i.e.~renormalizable) as 
\begin{equation}
V_0=\frac{1}{2}m^2\phi^2_c+\frac{\lambda}{4!}\phi^4_c \,\,,
\label{eq:SMtree}
\end{equation}
where we are only considering even powers of $\phi$ assuming there is a global symmetry $\phi_c\to -\phi_c$.

\subsubsection{One-loop potential and beyond}
At one-loop the effective potential should be computed as the sum of all one-loop diagrams, with an arbitrary number of external legs with zero momenta. In particular the $n$-th diagram has $n$ propagators, $n$ vertices and $2n$ external legs, corresponding to the presence of $\lambda^n$. The $n$ propagators will contribute $i^n(p^2-m^2+i\epsilon)^{-n}$, the external lines will contribute $\phi_c^{2n}$ and each vertex a factor of $-i\lambda/2$, where the factor of $1/2$ comes the symmetry of the diagram under exchange of two external lines. Finally there is a global symmetry factor $\frac{1}{2n}$ where the $1/n$ factor comes from the symmetry of the diagram under the discrete group $\mathbb Z_n$ and $1/2$ from the symmetry of the diagram under reflection. Finally there is integration over internal loop momentum and an extra global factor of $i$ from the definition of the generating functional. Putting all together one gets, after the Wick rotation with $p^0=ip_E^0$ and $p^2=(p^0)^2-\vec p^{\,2}=p_E^2$~\cite{Coleman:1973jx}
\begin{equation}
V_1(\phi_c)=i \sum_{n=1}^\infty\int \frac{d^4p}{(2\pi)^4}\frac{1}{2n}\left[\frac{\lambda\phi_c^2/2}{p^2-m^2+i\epsilon} \right]^n=-\frac{i}{2}\int\frac{d^4p}{(2\pi)^4}\log\left[1-\frac{\lambda\phi_c^2/2}{p^2-m^2+i\epsilon}\right]=\frac{1}{2}\int\frac{d^4p_E}{(2\pi)^4}\log\left[ 1+\frac{\lambda\phi_c^2/2}{p_E^2+m^2} \right]\,.
\end{equation}
The last expression can be written as 
\begin{equation}
V_1(\phi_c)=\frac{1}{2}\int\frac{d^4p_E}{(2\pi)^4}\log\left[ p_E^2+m^2(\phi_c) \right] \,\,,
\label{eq:1loopscalar}
\end{equation}
where $m^2(\phi_c)=d^2 V_0(\phi_c)/d\phi_c^2$ is the scalar mass for the background value $\phi_c$ of the scalar field (not necessarily at the tree level vacuum), and a constant $\phi_c$-independent term has been neglected.

The expression (\ref{eq:1loopscalar}) can be easily generalized to the case of several scalars $\phi^a_c$ ($a=1,\dots N_s$) with a tree-level potential $V_0(\phi^a,\phi_b^\dagger)$ as
\begin{equation}
V_{1,s}(\phi_c^a)=\frac{1}{2}\textrm{Tr} \int\frac{d^4p_E}{(2\pi)^4}\log\left[ p_E^2+M^2_s(\phi_c) \right]\,,\quad \textrm{where}\quad \left(M_s^2(\phi)\right)^a_b=\frac{\partial^2 V_0}{\partial\phi^\dagger_a\partial\phi^b}\,.
\end{equation}

A similar calculation for the case of fermion fields $\psi^A$ described by the Lagrangian
\begin{equation}
\mathcal L=i\bar\phi_A\gamma\cdot \partial\psi^A-\bar\psi_A\left( M_f \right)^A_B \psi^B\,,\quad \textrm{with}\quad \left( M_f \right)^A_B(\phi_c^a)=\sum_{a=1}^{N_s}\Gamma^A_{Ba}\phi^a_c\,,
\label{eq:lagrangian_f}
\end{equation}
where we are considering that fermions get a mass from the mechanism of symmetry breaking, yields
\begin{equation}
V_{1,f}(\phi_c^a)=-2\lambda_f\frac{1}{2}\textrm{Tr} \int\frac{d^4p_E}{(2\pi)^4}\log\left[ p_E^2+M^2_f(\phi_c) \right],
\end{equation}
where $\lambda_f=1$ (2) for Weyl (Dirac) fermions is just counting the number of fermion degrees of freedom. 

Finally,  gauge bosons $A_\mu\equiv A_\mu^\alpha T^\alpha$ (where $(T^\alpha)^{\,\beta}_\gamma\equiv i f^{\alpha\beta\gamma}$ are the group generators in the adjoint representation and $f^{\alpha\beta\gamma}$ the group structure constants) interact with the scalar fields through the Lagrangian, 
\begin{equation}
\mathcal L=-\frac{1}{4}\textrm{Tr}F_{\mu\nu}F^{\mu\nu}+\frac{1}{2}\textrm{Tr}(D_\mu\phi)^\dagger D^\mu\phi\,,\quad \textrm{where}\quad F_{\mu\nu}^\alpha=\partial_\mu A_\nu^\alpha-\partial_\nu A_\mu^\alpha +g f^{\alpha\beta\gamma}A_\mu^\beta A_\nu^\gamma
\label{eq:lagrangian_A}
\end{equation}
$D_\mu=\partial_\mu-igA_\mu^\alpha t_\alpha$ is the covariant derivative of the gauge group with gauge coupling $g$, and $(t_\alpha)^a_b$ are group generators in the group representation of $\phi^a$.
In the Landau-gauge, which does not require ghost-compensating terms, the only vertex which contributes to one-loop is the quartic coupling given by
\begin{equation}
\mathcal L_m=\frac{1}{2}\left(M_A \right)^2_{\alpha\beta}A^\alpha_\mu A^{\mu\beta}\,,\quad \textrm{where} \quad \left(M_A \right)^2_{\alpha\beta}=g^2 \left( t^a_{\alpha c}\phi_a\right)^\dagger t^c_{\beta b}\phi^b \,,
\end{equation}
The final expression for the effective potential is given by
\begin{equation}
V_{1,A}(\phi_c^a)=3\frac{1}{2}\textrm{Tr} \int\frac{d^4p_E}{(2\pi)^4}\log\left[ p_E^2+M^2_A(\phi_c) \right]\,,
\end{equation}
where the pre-factor 3 is the number of degrees of freedom of a massive gauge boson.

Computing the effective potential by summing infinite series of Feynman diagrams at zero external momentum beyond  one-loop is a tough task. However using the formalism of Ref.~\cite{Jackiw:1974cv} the task is affordable. To do that it is convenient to define the theory in the broken phase, by means of the expansion $\phi(x)=\phi_c+h(x)$, where $\phi_c$ is the background field value and $h(x)$ the excitation such that $\langle h(x)\rangle =0$, and the action
\begin{equation}
\hat S[\phi_c;h(x)]\equiv S[\phi(x)]-S[\phi_c]-h \cdot \frac{\delta S[\phi_c]}{\delta\phi_c}\,,
\end{equation} 
where the second term makes the vacuum energy equal to zero and the third therm is there to cancel the tadpole of the shifted action. 

\begin{figure}[htb]
	\centering
	\includegraphics[width=0.75\textwidth]{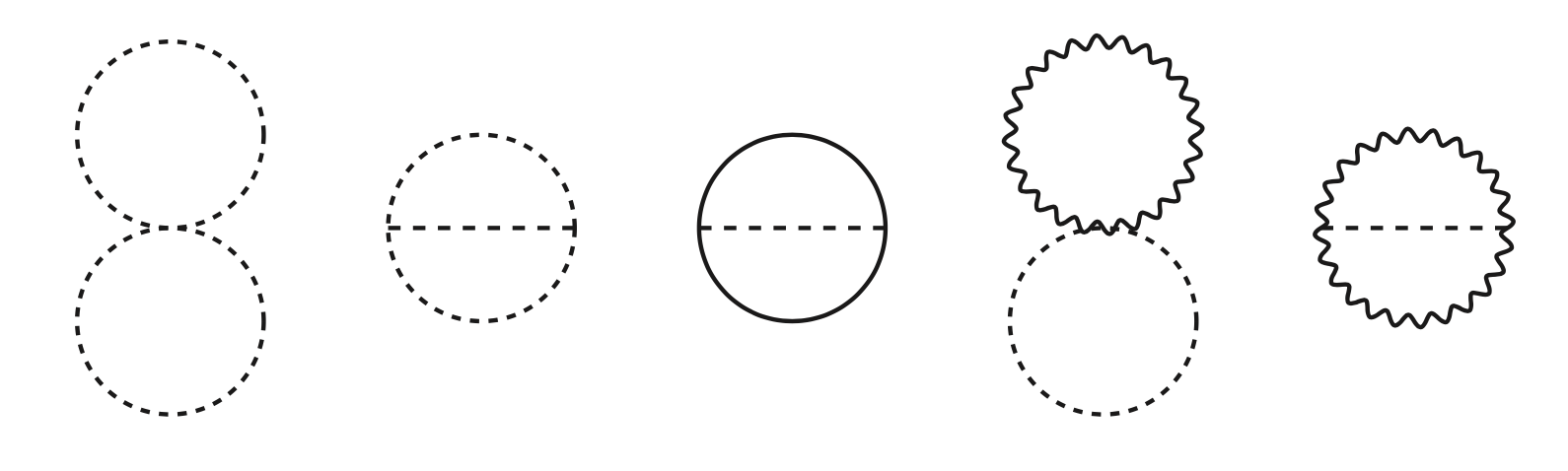}
	\caption{Two-loop diagrams contributing to the effective potential from scalar (dashed lines), fermion (solid lines) and gauge boson (wavy lines) fields. All propagators contain resummed $\phi_c$ dependent masses.}
	\label{fig:fig1}
\end{figure}

The corresponding shifted potential is given by
\begin{equation}
\hat V(\phi_c;h)=\frac{1}{2}m^2(\phi_c)h^2+\frac{\lambda}{3!}\phi_c h^3+\frac{\lambda}{4!}h^4 \,,
\end{equation}
which describes a theory for the field $h$ with shifted propagator $i(p^2-m^2(\phi_c)+i\epsilon)^{-1}$, with resummed mass $m^2(\phi_c)$, and vertices $h^3$ and $h^4$. The effective potential from scalar fields is described by vacuum diagrams and the one-loop result is trivially given by (\ref{eq:1loopscalar}) while there are 2 two-loop diagrams: \textit{i)} The eight-diagram, proportional to $\lambda$ and, \textit{ii)} The sunset diagram, proportional to $(\lambda\phi_c)^2$. In the same way using propagators with resummed masses the two-loop diagrams can be computed for fermion and gauge fields using the interactions in Eqs.~(\ref{eq:lagrangian_f}) and (\ref{eq:lagrangian_A}). The two loop diagrams are shown in Fig.~\ref{fig:fig1}.

For the relevant case of the Standard Model, one has a tree level potential in terms of the Higgs doublet $H\equiv (\chi_1+i\chi_2,\phi+i\chi_3)^T/\sqrt{2}$, with $\phi=\phi_c+h$, and mass spectrum given by
 %
%
\begin{equation}
V_0=-m^2|H|^2+\frac{\lambda}{6} |H|^4,\quad m_h^2(\phi_c)=\frac{\lambda}{2} \phi^2_c-m^2,\quad m_\chi^2(\phi_c)=\frac{\lambda}{6} \phi^2_c-m^2,\quad m_W^2(\phi_c)=\frac{g^2}{4}\phi^2_c,\quad m_Z^2(\phi_c)=\frac{g^2+g'^2}{4}\phi^2_c,\quad m_f^2(\phi_c)=\frac{h_f^2}{2}\phi^2_c \,,
\label{eq:SMmasses}
\end{equation}
where $g$ and $g'$ are the gauge couplings of SU(2) and U(1), respectively, and $h_f$ the fermion Yukawa coupling. 
Spontaneous breaking at a value $\phi_c=v\simeq 246$ GeV requires $m^2=\lambda v^2/6$, leading to massless Goldstone bosons at the minimum.

\subsubsection{Improved potential by the renormalization group equations}
The expressions for the effective potential we have deduced in the previous section are ultraviolet (UV) divergent. To make sense out of them we have to follow the renormalization procedure of quantum field theories. First of all we have to regularize the theory and all infinities must be absorbed by appropriate counterterms. The way the infinities are absorbed by the counterterms depend on the definition of the renormalized parameters, i.e.~on the choice on the renormalization conditions. Finally the theory written as a function of the renormalized parameters is finite.

The most convenient regularization scheme, which preserves the gauge invariance of the theory, is called dimensional regularization, and was introduced by t'Hooft and Veltman~\cite{tHooft:1972tcz}. It consists in making an analytic continuation of Feynman integrals to the complex plane in the number of space-time dimensions $4\to n=4-\epsilon$ so that the integrals have pole singularities in $(n-4)^{-1}$ which have to be subtracted out. Moreover, in working out the effective potential it is customary to use the so-called $\overline{\rm MS}$ renormalization scheme~\cite{tHooft:1973mfk} by which the quantity to be subtracted is $2/(4-n)-\gamma_E+\log(4\pi)$, where $\gamma_E=0.5772\dots$ is the Euler-Mascheroni constant. Then the effective potential (\ref{eq:1loopscalar}) can be written as
\begin{equation}
V_{1,s}(\phi_c)=\frac{1}{2}(\mu^2)^{2-n/2} \textrm{Tr}\int \frac{d^np}{(2\pi)^n}\log[p^2+M_s^2(\phi_c)] \,,
\end{equation}
where $\mu$ is a parameter with mass dimension which needs to be introduced to balance the dimension $n$ of the integration measure, and $p$ denotes the euclidean momentum. After subtracting the singularity in the limit $\epsilon\to 0$ the effective potential and the counterterms can be written as
\begin{equation}
V_{1,s}=\textrm{Tr}\frac{M_s^4(\phi_c)}{64\pi^2}\left[\log\frac{M_s^2(\phi_c)}{\mu^2}-\frac{3}{2}\right],\quad \delta\lambda=\frac{3\lambda^2}{32\pi^2}\left[\frac{2}{4-n}-\gamma_E+\log(4\pi) \right]\,.
\end{equation}
Similarly, for fermions and gauge bosons the one-loop renormalized contributions to the effective potential are given by
\begin{equation}
V_{1,f}=-2\lambda_f \textrm{Tr}\frac{M_f^4(\phi_c)}{64\pi^2}\left[\log\frac{M_f^2(\phi_c)}{\mu^2}-\frac{3}{2}\right],\quad V_{1,A}=3\,\textrm{Tr}\frac{M_A^4(\phi_c)}{64\pi^2}\left[\log\frac{M_A^2(\phi_c)}{\mu^2}-\frac{5}{6}\right]\,.
\label{eq:1loopfermion}
\end{equation}

Considering the previous results in this section and the SM mass spectrum, Eq.~(\ref{eq:SMmasses}), the one-loop SM effective potential is given in the Landau gauge as 
\begin{equation}
V^{\rm SM}_{1}(\phi_c)=\frac{1}{64\pi^2}\left\{m_h^4\left[\log\frac{m_h^2}{\mu^2}-\frac{3}{2} \right]+3m_\chi^4\left[\log\frac{m_\chi^2}{\mu^2}-\frac{3}{2} \right]+3m_Z^4\left[\log\frac{m_Z^2}{\mu^2}-\frac{5}{6} \right]+6m_W^4\left[\log\frac{m_W^2}{\mu^2}-\frac{5}{6} \right]-\sum_f 4 N_c m_f^4\left[\log\frac{m_f^2}{\mu^2}-\frac{3}{2} \right] \right\} 
\label{eq:SMpotential}
\end{equation} 
where the SM fermions are $f=u,d,c,s,t,b,e,\mu,\tau$ and $N_c=3\,(1)$ for quarks (leptons).

As we have seen above, the calculation of the effective action involves a scale $\mu$ which is unphysical in the sense that the theory should be independent of the chosen value of it. In fact a change on the value of $\mu$ should be accompanied by a change in the renormalized parameters $\lambda_i$ (coupling and masses) such that the theory remains unchanged. This requirement formally translates into the invariance equation
\begin{equation}
\frac{dV_{\rm eff}}{d\mu}=\left[\mu\frac{\partial}{\partial\mu}-\beta_i \frac{\partial}{\partial\lambda_i}-\gamma \phi_c \frac{\delta}{\delta \phi_c}  \right]V_{\rm eff}=0,
\label{eq:invariance}
\end{equation}
and $\gamma$ is the anomalous dimension of the scalar field. The formal solution to (\ref{eq:invariance}) is given by
\begin{equation}
V_{\rm eff}=V_{\rm eff}(\mu(t),\lambda_i(t),\phi(t)),\quad\textrm{where}\quad \mu(t)=\mu e^{t},\quad \phi(t)=\phi_c e^{-\int_0^t \gamma(\lambda_i(t'))dt'},\quad \beta_i=\frac{d\lambda_i(t)}{dt}\,.
\label{eq:Veffimp}
\end{equation}
In other words, in the so-called improved (by the renormalization group) effective potential at the $L$-loop order all masses and coupling run with the renormalization group at the $L+1$ loop order. The simplest approximation is where the effective potential is at the tree level ($L=0$), improved by the renormalization group at the $L=1$ order. The scale independence of the effective potential is achieved at all loop order. The more number of loops we consider the more the scale independence of the effective potential is achieved. 

\subsection{Effective potential at finite temperature: thermal corrections}

In the previous sections we have described the effective potential at zero temperature. However in early universe calculations particles live in a thermal plasma and the thermal corrections should be computed and taken into account. The thermal plasma is characterized by the temperature $T$ and chemical potentials $\mu_A$ associated with conserved global charges $Q_A$, and such that the density operator in the grand canonical ensemble is given by $\rho\propto \exp[\sum_A \mu_A/T-H/T]$, where $H$ is the theory hamiltonian, and normalized such that $\textrm{tr}\,\rho=1$.  In the following we will neglect the contribution from chemical potentials to the effective potential, as normally in the early universe $\mu_A\ll T$ (e.g.~the baryon number $\mu_B\ll T$).

In the imaginary time $\tau$ formalism, a thermal field theory can be well described by a theory defined on $S^1\times \mathbb R^3$ where the euclidean time is compactified on the circle $S^1$. As the time is compactified on a circle with radius $1/T$, where $T$ is the temperature, the energy is discretized and there appears the (tower of) Matsubara modes for which there is a mass component coming from the compactification process, on top of a possible four-dimensional invariant mass term. For reviews and textbooks see e.g.~Refs.~\cite{Landsman:1986uw,Quiros:1994dr,Das:1997gg,Quiros:1999jp,Bellac:2011kqa,Kraemmer:2003gd,Laine:2016hma}.

The effective potential computed from the sum of Matsubara modes, in the theory with compactified euclidean time, is finite as the theory has a natural cutoff at the value $\Lambda=T$. Bosonic (fermionic) fields are periodic (antiperiodic) under transformation $\tau\to \tau+\beta$, where $\beta=1/T$. The Matsubara modes for bosons (fermions) get at tree level, by the compactification procedure, discretized values of the energy $\omega_n=2n\pi/\beta$ ($\omega_n=(2n+1)\pi/\beta$) where $n=0,1,\dots$. Moreover the bosonic zero modes $n=0$ get one-loop square masses $(\propto g^2 T^2)$ from the interaction with the rest of fields. In the same way the complete effective potential can be computed using the Feynman rules of the thermal field theory~\cite{Quiros:1994dr}. 

We will skip the details of the calculation and just give the final one-loop result in terms of the thermal integrals for bosons $B$ with $n_B$ bosonic degrees of freedom, and fermions $F$ with $n_F$ fermionic degrees of freedom 
\begin{equation}
V_1(\phi_c;T)=\sum_B n_B \frac{T^4}{2\pi^2}J_B[m^2(\phi_c)/T^2]-2  \sum_F \lambda_F n_F \frac{T^4}{2\pi^2}J_F[m^2(\phi_c)/T^2],\quad \textrm{where}\quad J_{B,F}[m^2/T^2]=\int _0^\infty dx x^2 \log\left[ 1\mp e^{-\sqrt{x^2+m^2/T^2}} \right]\,.
\label{eq:potT}
\end{equation}
The thermal functions $J_{B,F}$ admit the high-temperature expansions for $m^2/T^2<1$ as
\begin{equation}
J_B(m^2/T^2)=-\frac{\pi^4}{45}+\frac{\pi^2}{12}\frac{m^2}{T^2}-\frac{\pi}{6}\left( \frac{m^2}{T^2}\right)^{3/2}-\frac{1}{32}\frac{m^4}{T^4}\log \frac{m^2}{a_B T^2}-2\pi^{7/2}
\sum_{\ell=1}^\infty
(-1)^\ell\frac{\zeta(2\ell+1)}{(\ell+1)!}\Gamma\left(\ell+\frac{1}{2} \right)\left( \frac{m^2}{4\pi^2 T^2} \right)^{\ell+2}
\label{eq:JB}
\end{equation}
and
\begin{equation}
J_F(m^2/T^2)=\frac{7\pi^4}{360}-\frac{\pi^2}{24}\frac{m^2}{T^2}-\frac{1}{32}\frac{m^4}{T^4}\log \frac{m^2}{a_F T^2}-\frac{\pi^{7/2}}{4}
\sum_{\ell=1}^\infty
(-1)^\ell\frac{\zeta(2\ell+1)}{(\ell+1)!} \left(1-2^{-2\ell-1} \right) \Gamma\left(\ell+\frac{1}{2} \right)\left( \frac{m^2}{4\pi^2 T^2} \right)^{\ell+2}\,,
\label{eq:JF}
\end{equation}
where $a_B=16\pi^2\exp(3/2-2\gamma_E)$, $a_F=\pi^2\exp(3/2-2\gamma_E)$, $\gamma_E$ is the Euler constant  and $\zeta$ is the Riemann $\zeta$-function. These expansions will be useful for determining the strength of cosmological phase transitions as we will see in the next section. We can already see that only bosons create a term in the expansion as $(m^2/T^2)^{3/2}$ which is responsible for a barrier at finite temperature between the symmetric and the broken phase and triggering thus a first order phase transition, relevant for creating the baryon-antibaryon asymmetry of the universe.

\section{Applications of the effective potential}
\label{sec:applications}
As we have already pointed out the effective potential is a useful tool to compute the vacuum (i.e.~the state of lowest energy) of a theory. The loop corrections to the effective potential are relevant to compute the vacuum of the theory, mainly in some specific cases, e.g.~radiative breaking, de Sitter expansion of the universe in an inflationary period, or the cosmological phase transitions. In this section we will illustrate some of the most relevant cases which have important physical applications.

\subsection{Radiative symmetry breaking}
\label{sec:radiative}
Given the shape of the one-loop effective potential, Eqs.~(\ref{eq:1loopscalar}) and (\ref{eq:1loopfermion}), one should fall into the mirage that one-loop corrections always trigger spontaneous symmetry breaking. Indeed this is not true. Let us first consider the simplest theory with a scalar field $\phi$ and a quartic potential $V_0=\frac{\lambda}{4!}\phi^4_c$ (symmetric under $\phi_c\to -\phi_c$) and the mass $m^2(\phi_c)=\lambda \phi_c^2/2$. The one-loop potential and the minimum condition for $\langle\phi_c\rangle$ read as
\begin{equation}
V(\phi_c)=\frac{\lambda}{4!}\phi^4_c+\frac{\lambda^2\phi_c^4}{256\pi^2}\log\left(\frac{\lambda \phi_c^2}{2\mu^2}-\frac{3}{2} \right) \,\quad \Rightarrow\quad  \lambda \log\frac{\lambda\langle\phi_c\rangle^2}{2\mu^2}-\lambda=-\frac{32\pi^2}{3}\,,
\label{eq:minimumfake}
\end{equation}
which, apparently, has a solution. Unfortunately the solution is outside the domain of applicability of perturbation theory, which requires that both $|\lambda|$ and $\log\frac{m^2(\phi_c)}{\mu^2}$ be small, a condition that is inconsistent with that implied by the last equation in (\ref{eq:minimumfake}). In this example the validity of perturbation theory prevents spontaneous breaking by loop corrections, a.k.a. radiative breaking. Is this the end of the story? Of course it is not. True radiative breaking can be implemented whenever loop corrections are smaller (or of the order of) the tree-level potential. 

A typical example, presented by Coleman and Weinberg in Ref.~\cite{Coleman:1973jx}, is massless scalar electrodynamics of two scalar (or one complex) fields $\phi_1$ and $\phi_2$ with charges $e$ and $-e$, respectively. The one-loop potential and its minimum are
\begin{equation}
V(\phi_c)=\frac{\lambda}{4!}\phi_c^4+\frac{\lambda^2\phi_c^4}{128\pi^2}\left(\log\frac{\lambda\phi_c^2}{2\mu^2}-\frac{3}{2}\right)+3\frac{e^4\phi_c^4}{64\pi^2}\left(\log\frac{e^2\phi_c^2}{\mu^2}-\frac{5}{6}  \right)\,\quad\Rightarrow\quad \lambda+\frac{3}{16\pi^2}\lambda^2\left(\log\frac{\lambda}{2e^2}-1 \right)=\frac{6e^4}{16\pi^2}\,,
\label{eq:minimum}
\end{equation}
where $\phi^2_c=\phi_1^2+\phi_2^2$ and we are fixing the scale $\mu=e\langle\phi_c\rangle$. Unlike in (\ref{eq:minimumfake}), the last equation in (\ref{eq:minimum}) can be easily satisfied if we choose $\lambda\sim e^4\ll 1$, so that the second term in Eq.~(\ref{eq:minimum}) can be neglected, as it is $\sim e^8$, leading to the value of $\lambda$ and the effective potential given by
\begin{equation}
\lambda\simeq \frac{6e^4}{16\pi^2}\quad \Rightarrow\quad V(\phi_c)\simeq 3 \frac{e^4\phi_c^4}{64\pi^2}\left(\log\frac{\phi_c^2}{\langle\phi_c\rangle^2} -\frac{1}{2} \right)
\end{equation}
without any reference to the parameter $\lambda$. In this way we originally have two dimensionless parameters $\lambda$ and $e$, and we end up with one dimensionless $e$ and one dimensionful  $\langle\phi_c\rangle$ parameters. The appearance of one dimensional parameter, out of a dimensionless one, sometimes is called \textit{dimensional transmutation}.

Another example of radiative breaking is when we consider the effective potential improved by the renormalization group equations. In this case the tree-level potential parameter $\lambda$, and the field $\phi$, are running with the renormalization group and so are $\mu$ dependent. If there are fermions in the theory, which contribute to the renormalization of the parameter $\lambda$ with a negative sign, it can happen that $\lambda(\mu_c)=0$ for some scale $\mu_c$ (see e.g.~Ref.~\cite{Buttazzo:2013uya}). In this case, for $\mu=\mu_c$ the effective potential in entirely dominated by loop corrections and it can trigger radiative breaking if the shape of the potential is such that it has a stable minimum. This case happens in the Standard Model, where the vanishing of the quartic coupling is triggered by the presence of the top-quark Yukawa coupling. It gives rise to a dangerous minimum which can destabilize the electroweak one, and will be the subject of the next section.

\subsection{Metastability of the Standard Model electroweak vacuum}
\label{sec:metastability}

In order to extrapolate the SM Higgs effective potential of Eq.~(\ref{eq:Veffimp}) at high values of the renormalization scale $\mu(t)$, one has to first assign the matching conditions for the SM running parameters, generically denoted by $\lambda_i(t)$, at some chosen reference scale, $\mu$. 
Here we exploit the matching conditions for the running parameters in the $\overline{\rm MS}$ renormalization scheme at $\mu=200$ GeV, which were derived in Ref.~\cite{Alam:2022cdv}. 
It is also useful to introduce the parameter $\alpha$, such that $\mu(t) =\alpha \phi(t)$.
Increasing levels of approximation for the Higgs effective potential as a function of $\phi(t)$ can thus be obtained, which in turn display an increasing level of independence on $\alpha$, see e.g.~Ref.~\cite{Casas:1996aq}.   

\begin{figure}[htb]
	\centering
	\includegraphics[width=0.475 \textwidth]{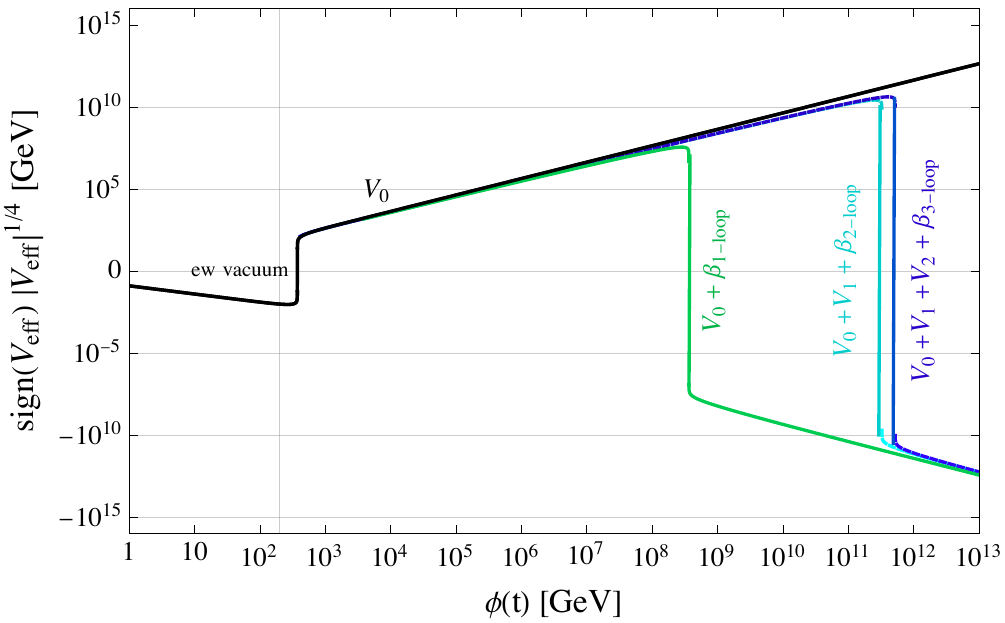}  \,\,\,\,\,\, \,\,\,\includegraphics[width=0.46 \textwidth]{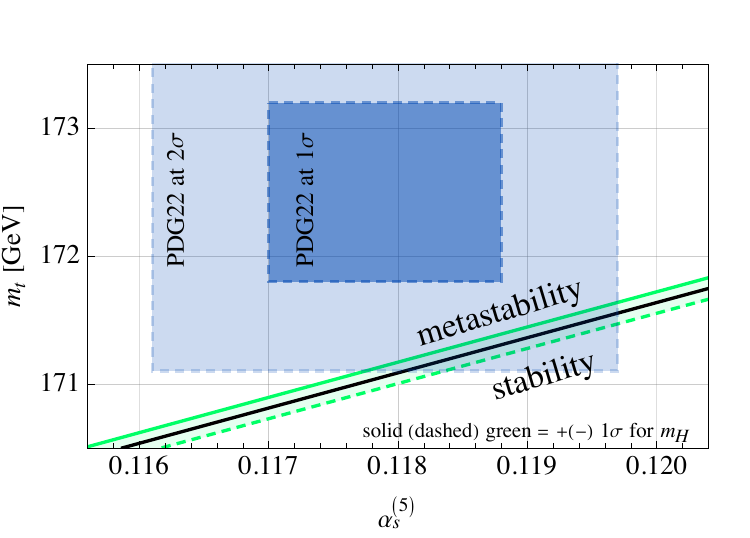}
	\caption{Left panel: Various levels of approximation for the Higgs effective potential, taking $\alpha=1$ and the PDG22 central values for the most relevant input parameters: 
	$m_t=172.5\pm 0.7$ GeV, $\alpha_s^{(5)}=0.1179\pm 0.0009$, $m_H=125.25\pm 0.17$ GeV~\cite{Workman:2022ynf}. 
	Right panel: the straight black line separating the regions where the electroweak vacuum is stable or metastable; 
	the close green solid and dashed lines display the effect of a positive and negative $1\,\sigma$ variation in $m_H$, 
	while the shaded blue rectangles display the $1\,\sigma$ and $2\,\sigma$ allowed ranges for $\alpha_s^{(5)}$ and $m_t$, according to Ref.~\cite{Workman:2022ynf}.
}
	\label{fig-meta}
\end{figure}

In order to illustrate the different loop contributions to the effective potential, the left plot of Fig.~\ref{fig-meta} displays the shape of the SM Higgs effective potential as a function of $\phi(t)$, for various levels of approximation, taking $\alpha=1$ and central values for the most relevant experimental inputs (mainly the top-quark and Higgs boson masses and the strong coupling constant), as given in Ref.~\cite{Workman:2022ynf}; due to the high accuracy achieved in the Higgs mass, the most relevant experimental inputs are the top quark mass and the strong coupling. The upper (black) line shows the tree-level result (also dubbed Mexican hat potential), $V_0(\phi)$ as given by Eq.~(\ref{eq:SMmasses}); at $\phi$ values larger than the electroweak minimum, $V_0(\phi)$ is an ever increasing function. As already mentioned, the tree-level one-loop improved Higgs potential, here denoted by $V_0+\beta_{1-loop}$, instead displays another minimum at high energy, deeper than the electroweak one, as an effect of the presence of the top-quark Yukawa coupling $h_t$ in the running of $\lambda(t)$~\cite{Cabibbo:1979ay}
\begin{equation}
\beta_\lambda^{\rm 1-loop}=\frac{d\lambda}{dt}=-\frac{36}{16\pi^2}h_t^4+\dots\quad \Rightarrow \quad \lambda(\phi)\simeq\lambda-\frac{36}{16\pi^2}h_t^4\log\frac{\phi}{v}\,
\label{eq:oneloop}
\end{equation}
where we are neglecting all couplings but the top Yukawa coupling. As we can see in Eq.~(\ref{eq:oneloop}) there is a minus sign in the $\beta$-function due to the presence of fermions in the loop and the RGE improved quartic coupling $\lambda$ becomes negative for field values much larger than the electroweak scale $\phi\gg v$. The full numerical result is shown by the solid green line in  the left panel of Fig.~\ref{fig-meta}, where the potential becomes negative at the field value $\phi(t)\sim 10^9$ GeV.
 
Moreover, the one-loop (or NLO) potential improved by the two-loop RGE, denoted by $V_0 + V_1+\beta_{2-loop}$, becomes negative at slightly larger field values, $\phi(t) \sim 10^{11}$ GeV; 
furthermore, the two-loop (or NNLO) result improved by the three-loop RGE, denoted by $V_0 + V_1+V_2+\beta_{3-loop}$, turns out to be nearly $\alpha$ independent and, consistently, it is close to the one-loop improved potential.

A metastable electroweak vacuum is thus suggested by the present central values of the relevant input parameters, $m_t$ and $\alpha_s^{(5)}$~\cite{Workman:2022ynf};
however, stability is still allowed within $2\,\sigma$, and it will be very difficult to discriminate between the two scenarios in the future~\cite{Franceschini:2022veh}. 
The right plot of Fig.~\ref{fig-meta} shows the regions corresponding to stability and metastability according to the two-loop improved Higgs potential, $V_0 + V_1+V_2(\phi)+\beta_{3-loop}$,
as a function of $\alpha_s^{(5)}$ and $m_t$; 
the shaded (green) band corresponds to a $1\,\sigma$ variation in the Higgs mass; the shaded (blue) rectangles correspond to the $1\,\sigma$ and $2\,\sigma$ allowed regions 
for $\alpha_s^{(5)}$ and $m_t$, according to Ref.~\cite{Workman:2022ynf}.

\begin{figure}[h]
	\centering
	\includegraphics[width=0.48 \textwidth]{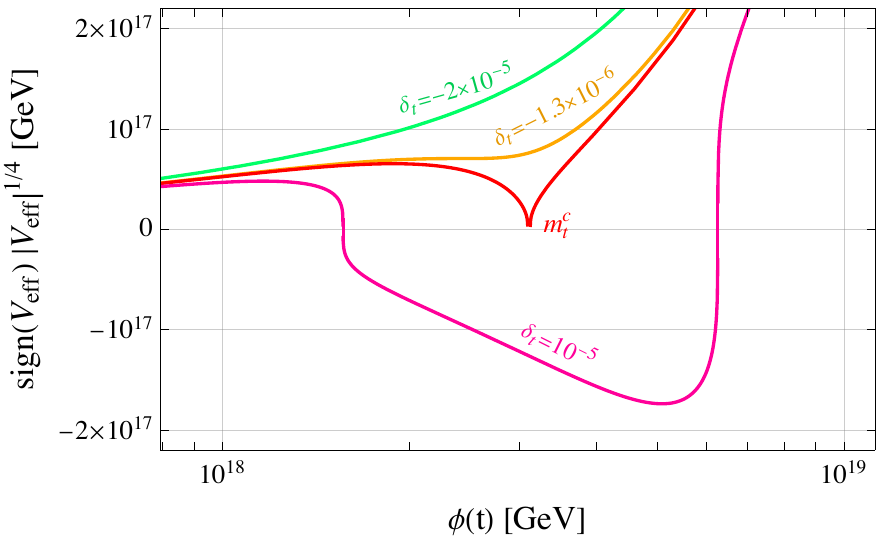}  \,\,\, \includegraphics[width=0.48 \textwidth]{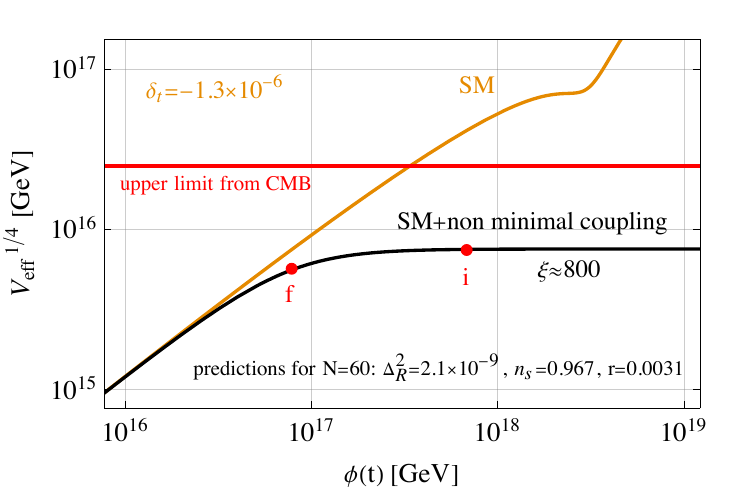}  
		\caption{Left panel: NNLO Higgs potential configurations for various values of the top mass near criticality. Right panel: The inflection point configuration for the Higgs effective potential in the Standard Model (orange curve). Including a non-minimal coupling to gravity with $\xi \approx 800$, the potential is flattened (black curve) and the cosmological predictions for $\Delta_R^2$, $n_s$ and $r$ are in agreement with 
		CMB data~\cite{Workman:2022ynf}. The red dots signal the beginning (i) and end (f) of inflation.}
	\label{fig-Hinfl}
\end{figure}

Taking central values for $\alpha_s^{(5)}$ and $m_H$~\cite{Workman:2022ynf}, the value of the top mass for which the NNLO Higgs effective potential displays two degenerate minima is $m_t^c \approx 171.0588$ GeV (see for instance Ref.~\cite{Masina:2024ybn}).  The shape of such critical configuration is shown (in red) in the left panel of Fig.\,\ref{fig-Hinfl}, where one can see that the high energy vacuum is located at field values $\phi(t)$ which are close to the Planck energy scale.  
It is useful to define the parameter $\delta_t= m_t/m_t^c -1$, and explore possible shapes close to the critical one, as for instance the inflection point configuration (in orange), achieved with $\delta_t = -1.3 \times 10^{-6}$, and the configurations corresponding to $\delta_t=-2\times 10^{-5}$ and $\delta_t=10^{-5}$.

\subsection{Quantum effects in Higgs inflation}
\label{sec:quantum}

The study of the shape of the Higgs effective potential at high energy is a relevant issue in order to understand whether the Higgs field might have played the role of the inflaton in the early universe. This issue has been widely discussed in the literature~\cite{Bezrukov:2007ep,Bezrukov:2010jz,Rubio:2018ogq}. For the sake of brevity, here we focus on slow-roll single-field inflation models~\cite{Liddle:1994dx}. 
A region where the Higgs potential $V(\phi)$ becomes sufficiently flat, for large enough values of the Higgs field, to meet the slow-roll conditions and cosmological observables
\begin{equation}
\epsilon\equiv \frac{M_P^2}{2}\left(\frac{V'}{V} \right)^2\ll 1,\quad |\eta|\equiv M_P^2\frac{|V''|}{V}\ll 1,\quad \Delta_R^2\simeq \frac{V}{24\pi^2 M_P^4\epsilon}\simeq 2.1\times 10^{-9},\quad n_s\simeq 1+2\eta-4\epsilon=0.9649(42),\quad r=16\epsilon<0.036
\end{equation}
is thus required, where $\Delta_R^2$ is the amplitude of density perturbations, $n_s$ the inflaton spectral index, $r$ the tensor-to-scalar ratio, and $M_P$ is the reduced Planck mass, $M_{P}=1/(8\pi G_{N})^{1/2}\simeq 2.43 \times 10^{18}$ GeV, 
where $G_N$ is the Newton constant. 

As previously discussed, the SM Higgs effective potential is compatible with an inflection point configuration. This scenario cannot however account for the observed cosmological
parameters~\cite{Masina:2018ejw}: as shown in the right plot of Fig.~\ref{fig-Hinfl}, the value of the Higgs effective potential at the stationary point exceeds the upper limit $V_{eff}^{1/4} \lesssim 2.5 \times 10^{16}$ GeV~\cite{Masina:2024ybn}, coming from CMB data by combining the amplitude of density perturbations, $\Delta_R^2$, 
with the upper bound on the tensor-to-scalar ratio $r \lesssim 0.036$~\cite{Workman:2022ynf}. 


Extensions of the Standard Model are thus needed. 
An interesting possibility is to modify the Higgs potential by introducing a non-minimal gravitational coupling $\xi$ between the SM Higgs doublet $H$ and the Ricci scalar $R$~\cite{Bezrukov:2007ep}. 
The classical action for Higgs inflation is
\begin{equation}
\mathcal{S}=\int d^{4}x \, \sqrt{-g}\left[\mathcal{L}_{\rm SM}-\frac{M_P^{2}}{2} R -\xi\left| H\right|^{2} R\right]\,,
\label{eq-act}
\end{equation}
where $g$ is the determinant of the FLRW 
metric, and 
$\mathcal{L}_{\rm SM}$ is the SM Lagrangian.
The effect of the introduction of the non-minimal coupling to gravity, $\xi$, is to flatten the Higgs potential at field values larger than about $M_P/\sqrt{\xi}$, so that the Higgs might successfully play the role of the inflaton~\cite{Bezrukov:2007ep}. Thanks to the flattening mechanism, 
configurations that in the Standard Model would be stable or even slightly metastable, become viable for inflation~\cite{Masina:2024ybn}; this happens for both the metric and the Palatini formalisms of gravity~\cite{Palatini:1919ffw}.
For instance, as shown in the right plot of Fig.\,\ref{fig-Hinfl} for the metric formalism, for the same input parameters that would lead to an inflection point configuration in the Standard Model, the inclusion of a non-minimal coupling, $\xi \approx 800$, suitably flattens the Higgs effective potential, leading to viable cosmological predictions for $\Delta_R^2$, $r$ and the spectral index $n_s$~\cite{Workman:2022ynf}. The much  debated issue of unitarity seems to not be problematic for the latter configuration: as the unitarity cutoff at large field values is $\Lambda \sim M_P/\sqrt{\xi} = 8.5 \times 10^{16}$ GeV,  the inflationary dynamics is expected to be reliable~\cite{Masina:2024ybn}.

\subsection{Cosmological phase transitions}
\label{sec:cosmological}
In the previous sections we covered applications of the effective potential methods in QFT at zero temperature. However in the early universe, as already commented, particles are propagating in a thermal plasma characterized by the temperature $T$ so that the potential depends, on top of the background field configuration $\phi$, on $T$ i.e.~$V(\phi,t)$. This makes it that the potential can have different minima at different temperatures, i.e.~different phases, and the balance in the free energy makes it possible the transition between them, what is called as \textit{phase transitions}.

Phase transitions are mainly classified into first and second order phase transitions depending on the shape of the potentials at different temperatures. Using the high temperature behavior from
(\ref{eq:potT}) one can typically describe the potential of a first order phase transition by~\cite{Dine:1992wr}
\begin{equation}
V(\phi,T)=\frac{1}{2}m^2\phi^2+\frac{\lambda}{4!}\phi^4+2 B v^2\phi^2-\frac{3}{2}B\phi^4+B\phi^4\log\frac{\phi^2}{v^2}+V_1(\phi,T) \simeq D(T^2-T_0^0)\phi^2-ET\phi^3+\frac{\lambda(T)}{4!}\phi^4 \,,
\label{eq:potT-bis}
\end{equation} 
where $B$, $D$, $T_0$ and $E$ are $T$-independent coefficients and $\lambda$ a slowly varying $T$-dependent function, see e.g.~Ref.~\cite{Quiros:1999jp}, where on-shell renormalization conditions  $V'_1(v)=V''_1(v) =0$ are imposed. Notice that at values $T\gg T_0$ the potential has a minimum at the origin $\phi=0$. At some temperature $T_1$ a local minimum first appears as an inflection point at a field value $\phi(T_1)\neq 0$, given by~\cite{Dine:1992vs}
\begin{equation}
T_1^2=\frac{4\lambda(T_1)DT_0^2}{4\lambda(T_1)D- 27\, E^2},\quad \phi(T_1)=\frac{9 E T_1}{\lambda(T_1)} \,,
\end{equation}
while a barrier between the latter and the origin starts to develop at lower temperatures and the inflection point splits into a maximum $\phi_M(T)$ and a local minimum $\phi_m(T)$ given by
\begin{equation}
\phi_M(T)=\frac{9ET}{\lambda(T)}-\frac{9}{\lambda(T)}\sqrt{E^2T^2-\frac{4 }{27}\lambda(T)D(T^2-T_0^2)},\quad 
\phi_m(T)=\frac{9ET}{\lambda(T)}+\frac{9}{\lambda(T)}\sqrt{E^2T^2-\frac{4}{27}\lambda(T)D(T^2-T_0^2)}\,.
\end{equation}
At the critical temperature $T_c$ the origin and the local minimum become degenerate, such that
\begin{equation}
T_c^2=\frac{\lambda(T_c)DT_0^2}{\lambda(T_c)D-6 E^2},\quad 
\phi_M(T_c)=\frac{6 ET_c}{\lambda(T_c)},\quad 
\phi_m(T_c)=\frac{12 ET_c}{\lambda(T_c)} \,.
\end{equation}

For $T<T_c$ the minimum at $\phi=0$ becomes metastable and the minimum at $\phi_m(T)$ becomes the global one. The phase transition starts at the nucleation temperature, $T_n<T_c$ by tunneling, which is more or less efficient depending on the heigh of the barrier and the shallowness of the potential. Even if the tunneling is very inefficient, at $T=T_0$ the barrier disappears, the origin becomes a maximum and the second minimum becomes equal to $\phi_m(T_0)=18ET_0/\lambda(T_0)$. 
As we can see the barrier is generated by the term proportional to $E$ in the effective potential. A quick glance at the third term of the expansion in Eq.~(\ref{eq:JB}) shows that the cubic term, and thus the presence of the barrier, is triggered by the propagation of bosons (in particular the zero modes of bosons) and not by the fermions, see Eq.~(\ref{eq:JF}), for which the Matsubara modes do not have any zero mode, see Ref.~\cite{Quiros:1999jp}.  

A first order phase transition proceeds abruptly either by thermal tunneling or by quantum tunneling, with a probability of decay of the false vacuum per unit time per unit volume given by
%
$\Gamma/\nu=Ae^{-B}$
%
where the pre-factor and exponent are $A\sim T^4$ and $B=S_3(T)/T$ for thermal tunneling, and $A\sim v^4$ and $B=S_4(T)$, where $v$ is the typical scale of the theory, for a quantum tunneling. 
$S_3(T)$ and $S_4(T)$ are the euclidean actions evaluated at the bounce solutions $S_E$ with symmetry $O(3)$ and $O(4)$, respectively, given by~\cite{Megias:2021rgh}
\begin{equation}
S_E(T)=\Omega_n \int d\sigma \sigma^{n-1}\left[ \frac{1}{2}\left(\frac{d\phi_B(\sigma)}{d\sigma} \right)^2 +V(\phi_B,T)\right], \ \Omega_n=n\pi^{n/2}/\Gamma(1+n/2),\ \textrm{with}\quad \left(n=3,\  \sigma=\sqrt{\vec x^{\;2}}\right)\quad \textrm{or}\quad \left(n=4,\ \sigma=\sqrt{\vec x^{\;2}+\tau^2}\right)\,.
\end{equation}
The bounce solutions are subject to the equations of motion and boundary conditions:
\begin{equation}
\frac{d^2 \phi_B(\sigma)}{d\sigma^2}+\frac{n-1}{\sigma}\frac{d\phi_B}{d\sigma}=\frac{\partial V}{\partial\phi},\quad \left.\frac{d\phi_B}{d\sigma}\right|_{\sigma=0}=0,\quad
\lim_{\sigma\to\infty}\phi_B=0\,.
\end{equation}
The bounce equation has to be solved numerically, although in the case we are considering here, for one simple field, it can be easily solved by the overshoot/undershoot method, although often analytical approximations, in particular thin or thick wall approximations, can be used.

Phase transitions then proceed at the so-called nucleation temperature $T_n$ such that there is one bubble of the broken phase per Hubble patch, and are completed at the percolation temperature $T_p$ such that the probability of any spatial point to be in the true minimum is larger than $30\%$~\cite{Ellis:2018mja}. In particular it is customary to parametrize the strength of the phase transition by the parameters 
\begin{equation}
\alpha(T)\equiv \frac{\Delta V}{\rho(T_n)},\quad \frac{\beta}{H_n}\equiv \left. T\frac{dS_E}{dT}\right|_{T=T_n}\,,
\end{equation}
where $\Delta V$ is the quantum or thermal jump in the potential, $\rho(T_n)$ is the radiation energy density in the symmetric phase and $\beta/H_n$ is the inverse duration of the phase transition. Typically very strong and supercooled phase transitions are characterized by $\alpha(T_n)\gg 1$ and $\beta/H_n\lesssim 10$. Under these conditions the first order phase transitions produce a stochastic gravitational waves background with an amplitude $h^2\bar\Omega_{\rm GW}$ and peak frequency $f_p$~\cite{Caprini:2024hue}, and primordial black holes with a mass $M_{\rm PBH}$~\cite{Liu:2021svg,Gouttenoire:2023naa,Conaci:2024tlc}, typically given by
\begin{equation}
h^2\bar\Omega_{\rm GW}\simeq 8\times 10^{-7}(H_n/\beta)^2,\quad \frac{f_p}{\rm Hz}\simeq 1.8\times 10^{-5}(\beta/H_n)\, (\textrm{TeV}/T_R)^2,\quad \frac{M_{\rm PBH}}{M_\odot}\simeq 9\times 10^{-9}(\textrm{TeV}/T_R)^2\,,
\end{equation}
where $T_R$ is the reheat temperature after the phase transition, and $M_\odot$ the solar mass.

For the Standard Model the parameters of the expansion in Eq.~(\ref{eq:potT-bis}) are given by~\cite{Dine:1992vs}
\begin{align}
D&\simeq\frac{2m_W^2+m_Z^2+2m_t^2}{8v^2},\quad E\simeq\frac{2m_W^3+m_Z^3}{4\pi v^3},\quad T_0^2\simeq \frac{m_H^2-8Bv^2}{4D},\quad B=3\frac{2m_W^4+m_Z^4-4m_t^4}{64\pi^2v^4},\nonumber\\
\lambda(T)&\simeq \lambda-\frac{9}{8\pi^2 v^4}\left(2m_W^4\log\frac{m_W^2}{A_B T^2}+m_Z^4\log\frac{m_Z^2}{A_B T^2}-4m_t^4\log\frac{m_t^2}{A_F T^2}  \right)\,,
\end{align}
where $\log A_B\simeq 3.9076$ and $\log A_F\simeq1.1351$. 

\begin{figure}[htb]
	\centering
	\includegraphics[width=0.45 \textwidth]{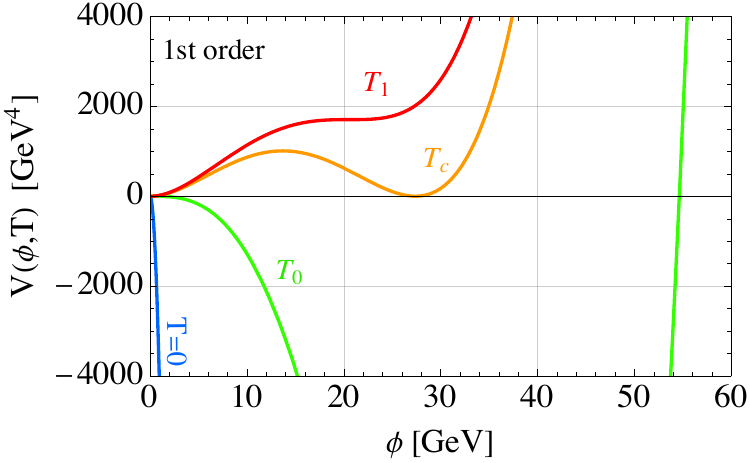} \,\,\,\,\,\,\,\,
	  \,\,\, \includegraphics[width=0.44 \textwidth]{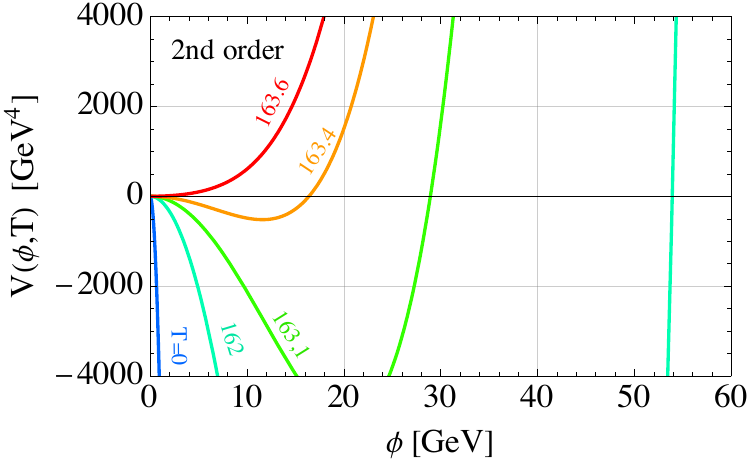}  
		\caption{Effective potential (normalized to zero at $\phi=0$). Left: for a first order phase transition, with $T=T_0\simeq 163.54$ GeV (green line), $T_c \simeq 163.94$ GeV (orange line), $T_1 \simeq 163.97$ GeV (red line). Right: for a second order phase transition, with the indicated values of the temperature, in GeV units.}
	\label{fig:phasetransition}
\end{figure}
A plot of the SM effective potential in the one-loop approximation is shown in the left panel of Fig.~\ref{fig:phasetransition} for different temperatures. As we can see the phase transition from Fig.~\ref{fig:phasetransition} is a weak first order phase transition, and $\phi(T_c)/T_c\simeq 0.17$, $\phi(T_0)/T_0\simeq 0.25$. In fact this is an artifact of the one-loop approximation and moreover of perturbation theory, and we show it here just for pedagogical purposes. In reality, non-perturbative lattice calculations show that the SM phase transition, for the experimental value of the Higgs mass, is a continuous crossover~\cite{Kajantie:1995kf}.

In general the phase transition for a theory whose high temperature expansion potential is given by Eq.~(\ref{eq:potT-bis}) with $E=0$ is called second order phase transition.
A second order phase transition is smooth as the curvature of the thermal potential is simply given by $m^2(\phi,T)=\lambda(T) \phi^2/2+2D(T^2-T_0^2)$ and the origin smoothly becomes a maximum at $T=T_0$ so that the minimum $\phi_m(T)$ evolves for $T<T_0$ with the temperature as
\begin{equation}
\phi_m(T)=2\sqrt{\frac{3D(T^2_0-T^2)}{\lambda(T)}} \quad \Rightarrow\quad  \phi_m(0)=2\sqrt{\frac{3D(T^2_0)}{\lambda(0)}},
\end{equation}
toward the zero temperature minimum. A second order phase transition smoothly proceeds by rolling down of the field from the origin toward the global minimum at $T\leq T_0$. For illustrative purposes, see the right panel of Fig.~\ref{fig:phasetransition}.

\section{Conclusions}
\label{sec:conclusions}
In this chapter we have presented some of the most basic results on the effective potential at zero temperature and in thermal field theories, which are of application in the early universe. At zero temperature the most relevant applications of the effective potential methods refer to the metastability of the electroweak minimum in the Standard Model and its relation to the measured values of the Higgs and top quark masses, and the strong coupling constant. This question is of upmost importance in view of the possibility of triggering cosmological inflation by the Higgs field, mainly when it is non-minimally coupled to gravity as inflation happens at scales higher than the instability scale of the electroweak minimum. A big question here, which attracts many people's attention, is how is it possible to avoid the metastability of the electroweak minimum and make it stable? This question is interesting in itself and is also related to the possibility of triggering cosmological inflation with the Standard Model fields. Also related to beyond the Standard Model physics.

A second line which has been covered by this chapter is the shape of cosmological phase transitions, induced by loop corrections in the thermal theory. This line is very relevant for future studies in beyond the Standard Model, as in the Standard Model the phase transition is very weak (actually it is a crossover as shown in non-perturbative/lattice calculations) a result implied by the small Standard Model value of the quartic coupling. One of the goals of present and future particle colliders is measuring the quartic coupling in processes with two or three Higgs particles in the final states. A departure from the Standard Model value could produce a strong first order phase transition, which could be detected by gravitational waves interferometers and thus signaling the presence of new physics beyond the Standard Model. The study of first order phase transitions, and the production of gravitational waves and primordial black holes, is thus a hot topic at this moment and worth of future studies. 

\begin{ack}[Acknowledgments]%
The work of MQ is supported by the grant PID2023-146686NB-C31 funded by MICIU/AEI/10.13039/501100011033/ and by FEDER, EU. 
IFAE is partially funded by the CERCA program of the Generalitat de Catalunya.
IM acknowledges partial support by the research project TAsP
(Theoretical Astroparticle Physics) funded by the Istituto Nazionale di Fisica Nucleare (INFN).
\end{ack}


\bibliographystyle{Numbered-Style} 
\bibliography{reference}

\begin{thebibliography*}{10}
\providecommand{\bibtype}[1]{}
\providecommand{\url}[1]{{\tt #1}}
\providecommand{\urlprefix}{URL }
\expandafter\ifx\csname urlstyle\endcsname\relax
  \providecommand{\doi}[1]{doi:\discretionary{}{}{}#1}\else
  \providecommand{\doi}{doi:\discretionary{}{}{}\begingroup
  \urlstyle{rm}\Url}\fi
\providecommand{\bibinfo}[2]{#2}
\providecommand{\eprint}[2][]{\url{#2}}
\makeatletter\def\@biblabel#1{\bibinfo{label}{[#1]}}\makeatother

\bibtype{Article}%
\bibitem{Heisenberg:1936nmg}
\bibinfo{author}{W. Heisenberg}, \bibinfo{author}{H. Euler},
  \bibinfo{title}{{Consequences of Dirac's theory of positrons}},
  \bibinfo{journal}{Z. Phys.} \bibinfo{volume}{98} (\bibinfo{number}{11-12})
  (\bibinfo{year}{1936}) \bibinfo{pages}{714--732},
  \bibinfo{doi}{\doi{10.1007/BF01343663}}, \eprint{physics/0605038}.

\bibtype{Article}%
\bibitem{Schwinger:1951nm}
\bibinfo{author}{Julian~S. Schwinger}, \bibinfo{title}{{On gauge invariance and
  vacuum polarization}}, \bibinfo{journal}{Phys. Rev.} \bibinfo{volume}{82}
  (\bibinfo{year}{1951}) \bibinfo{pages}{664--679},
  \bibinfo{doi}{\doi{10.1103/PhysRev.82.664}}.

\bibtype{Article}%
\bibitem{Goldstone:1962es}
\bibinfo{author}{Jeffrey Goldstone}, \bibinfo{author}{Abdus Salam},
  \bibinfo{author}{Steven Weinberg}, \bibinfo{title}{{Broken Symmetries}},
  \bibinfo{journal}{Phys. Rev.} \bibinfo{volume}{127} (\bibinfo{year}{1962})
  \bibinfo{pages}{965--970}, \bibinfo{doi}{\doi{10.1103/PhysRev.127.965}}.

\bibtype{Article}%
\bibitem{Jona-Lasinio:1964zvf}
\bibinfo{author}{G. Jona-Lasinio}, \bibinfo{title}{{Relativistic field theories
  with symmetry breaking solutions}}, \bibinfo{journal}{Nuovo Cim.}
  \bibinfo{volume}{34} (\bibinfo{year}{1964}) \bibinfo{pages}{1790--1795},
  \bibinfo{doi}{\doi{10.1007/BF02750573}}.

\bibtype{Article}%
\bibitem{ATLAS:2012yve}
\bibinfo{author}{Georges Aad}, et al. (\bibinfo{collaboration}{ATLAS}),
  \bibinfo{title}{{Observation of a new particle in the search for the Standard
  Model Higgs boson with the ATLAS detector at the LHC}},
  \bibinfo{journal}{Phys. Lett. B} \bibinfo{volume}{716} (\bibinfo{year}{2012})
  \bibinfo{pages}{1--29}, \bibinfo{doi}{\doi{10.1016/j.physletb.2012.08.020}},
  \eprint{1207.7214}.

\bibtype{Article}%
\bibitem{CMS:2012qbp}
\bibinfo{author}{Serguei Chatrchyan}, et al. (\bibinfo{collaboration}{CMS}),
  \bibinfo{title}{{Observation of a New Boson at a Mass of 125 GeV with the CMS
  Experiment at the LHC}}, \bibinfo{journal}{Phys. Lett. B}
  \bibinfo{volume}{716} (\bibinfo{year}{2012}) \bibinfo{pages}{30--61},
  \bibinfo{doi}{\doi{10.1016/j.physletb.2012.08.021}}, \eprint{1207.7235}.

\bibtype{Article}%
\bibitem{Coleman:1973jx}
\bibinfo{author}{Sidney~R. Coleman}, \bibinfo{author}{Erick~J. Weinberg},
  \bibinfo{title}{{Radiative Corrections as the Origin of Spontaneous Symmetry
  Breaking}}, \bibinfo{journal}{Phys. Rev. D} \bibinfo{volume}{7}
  (\bibinfo{year}{1973}) \bibinfo{pages}{1888--1910},
  \bibinfo{doi}{\doi{10.1103/PhysRevD.7.1888}}.

\bibtype{Article}%
\bibitem{Jackiw:1974cv}
\bibinfo{author}{R. Jackiw}, \bibinfo{title}{{Functional evaluation of the
  effective potential}}, \bibinfo{journal}{Phys. Rev. D} \bibinfo{volume}{9}
  (\bibinfo{year}{1974}) \bibinfo{pages}{1686},
  \bibinfo{doi}{\doi{10.1103/PhysRevD.9.1686}}.

\bibtype{Article}%
\bibitem{Iliopoulos:1974ur}
\bibinfo{author}{J. Iliopoulos}, \bibinfo{author}{C. Itzykson},
  \bibinfo{author}{Andre Martin}, \bibinfo{title}{{Functional Methods and
  Perturbation Theory}}, \bibinfo{journal}{Rev. Mod. Phys.}
  \bibinfo{volume}{47} (\bibinfo{year}{1975}) \bibinfo{pages}{165},
  \bibinfo{doi}{\doi{10.1103/RevModPhys.47.165}}.

\bibtype{Article}%
\bibitem{Martin:2001vx}
\bibinfo{author}{Stephen~P. Martin}, \bibinfo{title}{{Two Loop Effective
  Potential for a General Renormalizable Theory and Softly Broken
  Supersymmetry}}, \bibinfo{journal}{Phys. Rev. D} \bibinfo{volume}{65}
  (\bibinfo{year}{2002}) \bibinfo{pages}{116003},
  \bibinfo{doi}{\doi{10.1103/PhysRevD.65.116003}}, \eprint{hep-ph/0111209}.

\bibtype{Article}%
\bibitem{Martin:2015eia}
\bibinfo{author}{Stephen~P. Martin}, \bibinfo{title}{{Four-Loop Standard Model
  Effective Potential at Leading Order in QCD}}, \bibinfo{journal}{Phys. Rev.
  D} \bibinfo{volume}{92} (\bibinfo{number}{5}) (\bibinfo{year}{2015})
  \bibinfo{pages}{054029}, \bibinfo{doi}{\doi{10.1103/PhysRevD.92.054029}},
  \eprint{1508.00912}.

\bibtype{Article}%
\bibitem{Martin:2013gka}
\bibinfo{author}{Stephen~P. Martin}, \bibinfo{title}{{Three-Loop Standard Model
  Effective Potential at Leading Order in Strong and Top Yukawa Couplings}},
  \bibinfo{journal}{Phys. Rev. D} \bibinfo{volume}{89} (\bibinfo{number}{1})
  (\bibinfo{year}{2014}) \bibinfo{pages}{013003},
  \bibinfo{doi}{\doi{10.1103/PhysRevD.89.013003}}, \eprint{1310.7553}.

\bibtype{Article}%
\bibitem{Quiros:1994dr}
\bibinfo{author}{M. Quiros}, \bibinfo{title}{{Field theory at finite
  temperature and phase transitions}}, \bibinfo{journal}{Helv. Phys. Acta}
  \bibinfo{volume}{67} (\bibinfo{year}{1994}) \bibinfo{pages}{451--583}.

\bibtype{Inproceedings}%
\bibitem{Quiros:1999jp}
\bibinfo{author}{Mariano Quiros}, \bibinfo{title}{{Finite temperature field
  theory and phase transitions}}, in: \bibinfo{booktitle}{{ICTP Summer School
  in High-Energy Physics and Cosmology}} \bibinfo{year}{1999}, pp.
  \bibinfo{pages}{187--259}, \eprint{hep-ph/9901312}.

\bibtype{Article}%
\bibitem{Quiros:2007zz}
\bibinfo{author}{Mariano Quiros}, \bibinfo{title}{{Field theory at finite
  temperature and phase transitions}}, \bibinfo{journal}{Acta Phys. Polon. B}
  \bibinfo{volume}{38} (\bibinfo{year}{2007}) \bibinfo{pages}{3661--3703}.

\bibtype{Article}%
\bibitem{tHooft:1972tcz}
\bibinfo{author}{Gerard 't Hooft}, \bibinfo{author}{M.~J.~G. Veltman},
  \bibinfo{title}{{Regularization and Renormalization of Gauge Fields}},
  \bibinfo{journal}{Nucl. Phys. B} \bibinfo{volume}{44} (\bibinfo{year}{1972})
  \bibinfo{pages}{189--213}, \bibinfo{doi}{\doi{10.1016/0550-3213(72)90279-9}}.

\bibtype{Article}%
\bibitem{tHooft:1973mfk}
\bibinfo{author}{Gerard 't Hooft}, \bibinfo{title}{{Dimensional regularization
  and the renormalization group}}, \bibinfo{journal}{Nucl. Phys. B}
  \bibinfo{volume}{61} (\bibinfo{year}{1973}) \bibinfo{pages}{455--468},
  \bibinfo{doi}{\doi{10.1016/0550-3213(73)90376-3}}.

\bibtype{Article}%
\bibitem{Landsman:1986uw}
\bibinfo{author}{N.~P. Landsman}, \bibinfo{author}{C.~G. van Weert},
  \bibinfo{title}{{Real and Imaginary Time Field Theory at Finite Temperature
  and Density}}, \bibinfo{journal}{Phys. Rept.} \bibinfo{volume}{145}
  (\bibinfo{year}{1987}) \bibinfo{pages}{141},
  \bibinfo{doi}{\doi{10.1016/0370-1573(87)90121-9}}.

\bibtype{Book}%
\bibitem{Das:1997gg}
\bibinfo{author}{Ashok~K. Das}, \bibinfo{title}{{Finite Temperature Field
  Theory}}, \bibinfo{publisher}{World Scientific}, \bibinfo{address}{New York}
  \bibinfo{year}{1997}, ISBN \bibinfo{isbn}{978-981-02-2856-9,
  978-981-4498-23-4}.

\bibtype{Book}%
\bibitem{Bellac:2011kqa}
\bibinfo{author}{Michel~Le Bellac}, \bibinfo{title}{{Thermal Field Theory}},
  Cambridge Monographs on Mathematical Physics, \bibinfo{publisher}{Cambridge
  University Press} \bibinfo{year}{2011}, ISBN
  \bibinfo{isbn}{978-0-511-88506-8, 978-0-521-65477-7},
  \bibinfo{doi}{\doi{10.1017/CBO9780511721700}}.

\bibtype{Article}%
\bibitem{Kraemmer:2003gd}
\bibinfo{author}{Ulrike Kraemmer}, \bibinfo{author}{Anton Rebhan},
  \bibinfo{title}{{Advances in perturbative thermal field theory}},
  \bibinfo{journal}{Rept. Prog. Phys.} \bibinfo{volume}{67}
  (\bibinfo{year}{2004}) \bibinfo{pages}{351},
  \bibinfo{doi}{\doi{10.1088/0034-4885/67/3/R05}}, \eprint{hep-ph/0310337}.

\bibtype{Book}%
\bibitem{Laine:2016hma}
\bibinfo{author}{Mikko Laine}, \bibinfo{author}{Aleksi Vuorinen},
  \bibinfo{title}{{Basics of Thermal Field Theory}}, \bibinfo{comment}{vol.}
  \bibinfo{volume}{925}, \bibinfo{publisher}{Springer} \bibinfo{year}{2016},
  \bibinfo{doi}{\doi{10.1007/978-3-319-31933-9}}, \eprint{1701.01554}.

\bibtype{Article}%
\bibitem{Buttazzo:2013uya}
\bibinfo{author}{Dario Buttazzo}, \bibinfo{author}{Giuseppe Degrassi},
  \bibinfo{author}{Pier~Paolo Giardino}, \bibinfo{author}{Gian~F. Giudice},
  \bibinfo{author}{Filippo Sala}, \bibinfo{author}{Alberto Salvio},
  \bibinfo{author}{Alessandro Strumia}, \bibinfo{title}{{Investigating the
  near-criticality of the Higgs boson}}, \bibinfo{journal}{JHEP}
  \bibinfo{volume}{12} (\bibinfo{year}{2013}) \bibinfo{pages}{089},
  \bibinfo{doi}{\doi{10.1007/JHEP12(2013)089}}, \eprint{1307.3536}.

\bibtype{Article}%
\bibitem{Alam:2022cdv}
\bibinfo{author}{Zamiul Alam}, \bibinfo{author}{Stephen~P. Martin},
  \bibinfo{title}{{Standard model at 200~GeV}}, \bibinfo{journal}{Phys. Rev. D}
  \bibinfo{volume}{107} (\bibinfo{number}{1}) (\bibinfo{year}{2023})
  \bibinfo{pages}{013010}, \bibinfo{doi}{\doi{10.1103/PhysRevD.107.013010}},
  \eprint{2211.08576}.

\bibtype{Article}%
\bibitem{Casas:1996aq}
\bibinfo{author}{J.~A. Casas}, \bibinfo{author}{J.~R. Espinosa},
  \bibinfo{author}{M. Quiros}, \bibinfo{title}{{Standard model stability bounds
  for new physics within LHC reach}}, \bibinfo{journal}{Phys. Lett. B}
  \bibinfo{volume}{382} (\bibinfo{year}{1996}) \bibinfo{pages}{374--382},
  \bibinfo{doi}{\doi{10.1016/0370-2693(96)00682-X}}, \eprint{hep-ph/9603227}.

\bibtype{Article}%
\bibitem{Workman:2022ynf}
\bibinfo{author}{R.~L. Workman}, et al. (\bibinfo{collaboration}{Particle Data
  Group}), \bibinfo{title}{{Review of Particle Physics}},
  \bibinfo{journal}{PTEP} \bibinfo{volume}{2022} (\bibinfo{year}{2022})
  \bibinfo{pages}{083C01}, \bibinfo{doi}{\doi{10.1093/ptep/ptac097}}.

\bibtype{Article}%
\bibitem{Cabibbo:1979ay}
\bibinfo{author}{N. Cabibbo}, \bibinfo{author}{L. Maiani}, \bibinfo{author}{G.
  Parisi}, \bibinfo{author}{R. Petronzio}, \bibinfo{title}{{Bounds on the
  Fermions and Higgs Boson Masses in Grand Unified Theories}},
  \bibinfo{journal}{Nucl. Phys. B} \bibinfo{volume}{158} (\bibinfo{year}{1979})
  \bibinfo{pages}{295--305}, \bibinfo{doi}{\doi{10.1016/0550-3213(79)90167-6}}.

\bibtype{Article}%
\bibitem{Franceschini:2022veh}
\bibinfo{author}{Roberto Franceschini}, \bibinfo{author}{Roberto Franceschini},
  \bibinfo{author}{Alessandro Strumia}, \bibinfo{author}{Alessandro Strumia},
  \bibinfo{author}{Andrea Wulzer}, \bibinfo{author}{Andrea Wulzer},
  \bibinfo{title}{{The collider landscape: which collider for establishing the
  SM instability?}}, \bibinfo{journal}{JHEP} \bibinfo{volume}{08}
  (\bibinfo{year}{2022}) \bibinfo{pages}{229},
  \bibinfo{doi}{\doi{10.1007/JHEP08(2022)229}}, \bibinfo{note}{[Erratum: JHEP
  03, 167 (2023)]}, \eprint{2203.17197}.

\bibtype{Article}%
\bibitem{Masina:2024ybn}
\bibinfo{author}{Isabella Masina}, \bibinfo{author}{Mariano Quiros},
  \bibinfo{title}{{Electroweak metastability and Higgs inflation}},
  \bibinfo{journal}{Eur. Phys. J. C} \bibinfo{volume}{84}
  (\bibinfo{number}{11}) (\bibinfo{year}{2024}) \bibinfo{pages}{1153},
  \bibinfo{doi}{\doi{10.1140/epjc/s10052-024-13522-x}}, \eprint{2403.02461}.

\bibtype{Article}%
\bibitem{Bezrukov:2007ep}
\bibinfo{author}{Fedor~L. Bezrukov}, \bibinfo{author}{Mikhail Shaposhnikov},
  \bibinfo{title}{{The Standard Model Higgs boson as the inflaton}},
  \bibinfo{journal}{Phys. Lett. B} \bibinfo{volume}{659} (\bibinfo{year}{2008})
  \bibinfo{pages}{703--706},
  \bibinfo{doi}{\doi{10.1016/j.physletb.2007.11.072}}, \eprint{0710.3755}.

\bibtype{Article}%
\bibitem{Bezrukov:2010jz}
\bibinfo{author}{F. Bezrukov}, \bibinfo{author}{A. Magnin}, \bibinfo{author}{M.
  Shaposhnikov}, \bibinfo{author}{S. Sibiryakov}, \bibinfo{title}{{Higgs
  inflation: consistency and generalisations}}, \bibinfo{journal}{JHEP}
  \bibinfo{volume}{01} (\bibinfo{year}{2011}) \bibinfo{pages}{016},
  \bibinfo{doi}{\doi{10.1007/JHEP01(2011)016}}, \eprint{1008.5157}.

\bibtype{Article}%
\bibitem{Rubio:2018ogq}
\bibinfo{author}{Javier Rubio}, \bibinfo{title}{{Higgs inflation}},
  \bibinfo{journal}{Front. Astron. Space Sci.} \bibinfo{volume}{5}
  (\bibinfo{year}{2019}) \bibinfo{pages}{50},
  \bibinfo{doi}{\doi{10.3389/fspas.2018.00050}}, \eprint{1807.02376}.

\bibtype{Article}%
\bibitem{Liddle:1994dx}
\bibinfo{author}{Andrew~R. Liddle}, \bibinfo{author}{Paul Parsons},
  \bibinfo{author}{John~D. Barrow}, \bibinfo{title}{{Formalizing the slow roll
  approximation in inflation}}, \bibinfo{journal}{Phys. Rev. D}
  \bibinfo{volume}{50} (\bibinfo{year}{1994}) \bibinfo{pages}{7222--7232},
  \bibinfo{doi}{\doi{10.1103/PhysRevD.50.7222}}, \eprint{astro-ph/9408015}.

\bibtype{Article}%
\bibitem{Masina:2018ejw}
\bibinfo{author}{Isabella Masina}, \bibinfo{title}{{Ruling out Critical Higgs
  Inflation?}}, \bibinfo{journal}{Phys. Rev. D} \bibinfo{volume}{98}
  (\bibinfo{number}{4}) (\bibinfo{year}{2018}) \bibinfo{pages}{043536},
  \bibinfo{doi}{\doi{10.1103/PhysRevD.98.043536}}, \eprint{1805.02160}.

\bibtype{Article}%
\bibitem{Palatini:1919ffw}
\bibinfo{author}{Attilio Palatini}, \bibinfo{title}{{Deduzione invariantiva
  delle equazioni gravitazionali dal principio di Hamilton}},
  \bibinfo{journal}{Rend. Circ. Mat. Palermo} \bibinfo{volume}{43}
  (\bibinfo{number}{1}) (\bibinfo{year}{1919}) \bibinfo{pages}{203--212},
  \bibinfo{doi}{\doi{10.1007/BF03014670}}.

\bibtype{Article}%
\bibitem{Dine:1992wr}
\bibinfo{author}{Michael Dine}, \bibinfo{author}{Robert~G. Leigh},
  \bibinfo{author}{Patrick~Y. Huet}, \bibinfo{author}{Andrei~D. Linde},
  \bibinfo{author}{Dmitri~A. Linde}, \bibinfo{title}{{Towards the theory of the
  electroweak phase transition}}, \bibinfo{journal}{Phys. Rev. D}
  \bibinfo{volume}{46} (\bibinfo{year}{1992}) \bibinfo{pages}{550--571},
  \bibinfo{doi}{\doi{10.1103/PhysRevD.46.550}}, \eprint{hep-ph/9203203}.

\bibtype{Article}%
\bibitem{Dine:1992vs}
\bibinfo{author}{Michael Dine}, \bibinfo{author}{Robert~G. Leigh},
  \bibinfo{author}{Patrick Huet}, \bibinfo{author}{Andrei~D. Linde},
  \bibinfo{author}{Dmitri~A. Linde}, \bibinfo{title}{{Comments on the
  electroweak phase transition}}, \bibinfo{journal}{Phys. Lett. B}
  \bibinfo{volume}{283} (\bibinfo{year}{1992}) \bibinfo{pages}{319--325},
  \bibinfo{doi}{\doi{10.1016/0370-2693(92)90026-Z}}, \eprint{hep-ph/9203201}.

\bibtype{Article}%
\bibitem{Megias:2021rgh}
\bibinfo{author}{Eugenio Megias}, \bibinfo{author}{Germano Nardini},
  \bibinfo{author}{Mariano Quiros}, \bibinfo{title}{{Radion dynamics, heavy
  Kaluza\textendash{}Klein resonances and gravitational waves}},
  \bibinfo{journal}{Int. J. Mod. Phys. A} \bibinfo{volume}{37}
  (\bibinfo{number}{33}) (\bibinfo{year}{2022}) \bibinfo{pages}{2240023},
  \bibinfo{doi}{\doi{10.1142/S0217751X22400231}}, \eprint{2103.02705}.

\bibtype{Article}%
\bibitem{Ellis:2018mja}
\bibinfo{author}{John Ellis}, \bibinfo{author}{Marek Lewicki},
  \bibinfo{author}{Jos\'e~Miguel No}, \bibinfo{title}{{On the Maximal Strength
  of a First-Order Electroweak Phase Transition and its Gravitational Wave
  Signal}}, \bibinfo{journal}{JCAP} \bibinfo{volume}{04} (\bibinfo{year}{2019})
  \bibinfo{pages}{003}, \bibinfo{doi}{\doi{10.1088/1475-7516/2019/04/003}},
  \eprint{1809.08242}.

\bibtype{Article}%
\bibitem{Caprini:2024hue}
\bibinfo{author}{Chiara Caprini}, \bibinfo{author}{Ryusuke Jinno},
  \bibinfo{author}{Marek Lewicki}, \bibinfo{author}{Eric Madge},
  \bibinfo{author}{Marco Merchand}, \bibinfo{author}{Germano Nardini},
  \bibinfo{author}{Mauro Pieroni}, \bibinfo{author}{Alberto Roper~Pol},
  \bibinfo{author}{Ville Vaskonen} (\bibinfo{collaboration}{LISA Cosmology
  Working Group}), \bibinfo{title}{{Gravitational waves from first-order phase
  transitions in LISA: reconstruction pipeline and physics interpretation}},
  \bibinfo{journal}{JCAP} \bibinfo{volume}{10} (\bibinfo{year}{2024})
  \bibinfo{pages}{020}, \bibinfo{doi}{\doi{10.1088/1475-7516/2024/10/020}},
  \eprint{2403.03723}.

\bibtype{Article}%
\bibitem{Liu:2021svg}
\bibinfo{author}{Jing Liu}, \bibinfo{author}{Ligong Bian},
  \bibinfo{author}{Rong-Gen Cai}, \bibinfo{author}{Zong-Kuan Guo},
  \bibinfo{author}{Shao-Jiang Wang}, \bibinfo{title}{{Primordial black hole
  production during first-order phase transitions}}, \bibinfo{journal}{Phys.
  Rev. D} \bibinfo{volume}{105} (\bibinfo{number}{2}) (\bibinfo{year}{2022})
  \bibinfo{pages}{L021303}, \bibinfo{doi}{\doi{10.1103/PhysRevD.105.L021303}},
  \eprint{2106.05637}.

\bibtype{Article}%
\bibitem{Gouttenoire:2023naa}
\bibinfo{author}{Yann Gouttenoire}, \bibinfo{author}{Tomer Volansky},
  \bibinfo{title}{{Primordial black holes from supercooled phase transitions}},
  \bibinfo{journal}{Phys. Rev. D} \bibinfo{volume}{110} (\bibinfo{number}{4})
  (\bibinfo{year}{2024}) \bibinfo{pages}{043514},
  \bibinfo{doi}{\doi{10.1103/PhysRevD.110.043514}}, \eprint{2305.04942}.

\bibtype{Article}%
\bibitem{Conaci:2024tlc}
\bibinfo{author}{Angela Conaci}, \bibinfo{author}{Luigi Delle~Rose},
  \bibinfo{author}{P.~S.~Bhupal Dev}, \bibinfo{author}{Anish Ghoshal},
  \bibinfo{title}{{Slaying Axion-Like Particles via Gravitational Waves and
  Primordial Black Holes from Supercooled Phase Transition}}
  (\bibinfo{year}{2024}), \eprint{2401.09411}.

\bibtype{Article}%
\bibitem{Kajantie:1995kf}
\bibinfo{author}{K. Kajantie}, \bibinfo{author}{M. Laine}, \bibinfo{author}{K.
  Rummukainen}, \bibinfo{author}{Mikhail~E. Shaposhnikov}, \bibinfo{title}{{The
  Electroweak phase transition: A Nonperturbative analysis}},
  \bibinfo{journal}{Nucl. Phys. B} \bibinfo{volume}{466} (\bibinfo{year}{1996})
  \bibinfo{pages}{189--258}, \bibinfo{doi}{\doi{10.1016/0550-3213(96)00052-1}},
  \eprint{hep-lat/9510020}.

\end{thebibliography*}

\end{document}